\documentclass[12]{article}
\usepackage{amsmath,amssymb,amsthm,bm,blkarray}
\usepackage{mathtools}
\usepackage{mathrsfs}
\usepackage{bm}
\usepackage{slashed} 
\usepackage{multirow}
\usepackage{tikz}
\usepackage[utf8]{inputenc}
\usepackage{fullpage}
\usepackage{arydshln}
\usepackage{setspace}
\usepackage[margin=1in]{geometry}
\usepackage{dsfont}
\usepackage{graphicx }
\graphicspath{ {graphs/} }
\usepackage{subcaption}
\usepackage{float}
\usepackage{xcolor}
\usepackage{cite}
\usepackage{subcaption}
\usepackage{mwe}
\usepackage[toc,page]{appendix}
\usepackage{hyperref}
\usepackage{physics}
\hypersetup{colorlinks=true,linkcolor=blue,anchorcolor=blue,citecolor=green,filecolor=blue
urlcolor=blue,bookmarksnumbered=true,pdfview=FitB}

\theoremstyle{definition}

\usepackage{qcircuit}

\usepackage{filemod}

\usepackage{epstopdf}

\begin{document}

\begin{center}
{\Large \bf \strut
{\it Ab initio} nuclear structure via quantum adiabatic algorithm \\
\strut} 
{\large 
Weijie Du$^{a,b}$, James P. Vary$^a$, Xingbo Zhao$^{b,c}$, and Wei Zuo$^{b,c}$
}  

\vspace{5mm}
\noindent
{\small $^a$\it Department of Physics and Astronomy, Iowa State University, Ames, Iowa 50010, USA } \\
{\small $^b$\it Institute of Modern Physics, Chinese Academy of Sciences, Lanzhou 730000, China} \\
{\small $^c$\it School of Nuclear Science and Technology, University of Chinese Academy of Sciences, Beijing 100049, China} \\

\end{center}

\section*{Abstract}
\paragraph{Background:}
Solving nuclear many-body problems with an {\it ab initio} approach is widely recognized as a computationally challenging problem. Quantum computers offer a promising path to address this challenge. There are urgent needs to develop quantum algorithms for this purpose.  

\paragraph{Objective:}
In this work, we explore the application of the quantum algorithm of adiabatic state preparation with quantum phase estimation in {\it ab initio} nuclear structure theory. We focus on solving the low-lying spectra (including both the ground and excited states) of simple nuclear systems.

\paragraph{Ideas:}
The efficiency of this algorithm is hindered by the emergence of small energy gaps (level crossings) during the adiabatic evolution. In order to improve the efficiency, we introduce techniques to avoid level crossings: 1) by suitable design of the reference Hamiltonian; 2) by insertions of perturbation terms to modify the adiabatic path. 

\paragraph{Results:}
We illustrate this algorithm by solving the deuteron ground state energy and the spectrum of the deuteron bounded in a harmonic oscillator trap implementing the IBM Qiskit quantum simulator. The quantum results agree well the classical results obtained by matrix diagonalization.

\paragraph{Outlook:}
With our improvements to the efficiency, this algorithm provides a promising tool for investigating the low-lying spectra of complex nuclei on future quantum computers.  

\section{Introduction}

{\it Ab initio} nuclear structure theories are powerful tools for investigating nuclear properties \cite{Pieper:2001mp,Hagen:2013nca,Caurier:2004gf,Barrett:2013nh,Suhonen:2007}. One approach \cite{Barrett:2013nh} resembles configuration interaction (CI) methods in atomic physics and quantum chemistry since it seeks the Hamiltonian eigenvalues and eigenvectors in a chosen basis space. Such theories formulate the nuclear many-body problem in terms of the strong inter-nucleon interactions within a Hamiltonian that acts on all the nucleons in the nucleus. The three-dimensional harmonic oscillator (3DHO) basis is commonly adopted. Within the basis representation, the Hamiltonian of the nuclear system, incorporating the nucleon kinetic energies with the inter-nucleon interactions, generates a sparse matrix of large dimension \cite{Caurier:2004gf,Barrett:2013nh,Vary:2009qp,Maris:2008ax,Vary:2018hdv}. One of the main goals is to compute the low-lying spectrum to compare with and/or predict experimental results. 

However, the reach of the CI {\it ab initio} method (e.g., the no-core shell model \cite{Barrett:2013nh,Navratil:2000ww,Navratil:2000gs}) is restricted by available computational resources \cite{Vary:2009qp,Maris:2008ax,Vary:2018hdv}. Due to the complex nature of the nuclear many-body problem, the dimension of the model space grows exponentially with increasing number of constituent nucleons. The resources necessary for obtaining converged results with realistic inter-nucleon interactions exceed those available on world-leading supercomputers for all but nuclei of low atomic number, typically less than 20 nucleons. Quantum computers hold the promise of reducing the exponential scaling requirements to polynomial \cite{Feynman:1982}; and quantum computing techniques would open up a fruitful path to {\it ab initio} solutions for nuclei spanning the full table of nuclear isotopes.

For these reasons, there is a great need to develop quantum algorithms for performing nuclear structure calculation on quantum computers. Several heuristic and ground-breaking algorithms have been proposed in quantum chemistry and condensed matter physics \cite{Cao:2019,McArdle:2020}, including the adiabatic state preparation with quantum phase estimation (ASP-QPE) \cite{Aspuru-Guzik:2005,Whitfield:2011,Albash:2018}, the hybrid quantum-classical methods such as the variational quantum eigensolver (VQE) \cite{Peruzzo:2014,McClean:2016,Grimsley:2019}, the quantum approximate optimization algorithm \cite{Farhi:2014}, the quantum imaginary time evolution \cite{McArdle:2019xxx, Motta:2020} and the quantum Lanczos methods \cite{Motta:2020,Aydeniz:2020xxx}. Among these, the VQE method has been adopted to treat simple nuclear systems such as the deuteron \cite{Dumitrescu:2018,DiMatteo:2020dhe}. This successful hybrid method is flexible with respect to circuit depth and can be executed on near-term quantum computers \cite{Preskill:2018}. However, it is still unclear if this method can be easily adopted to excited states and to more complex nuclei (for one reason, the VQE method entails a judicious design of the trial function and additional noisy high-dimensional classical optimization \cite{McClean:2018}). 

In this work, we explore an alternative idea of using the ASP-QPE algorithm to calculate the low-lying spectrum of the nuclear system. In principle, if one starts with the ground state of the reference Hamiltonian and slowly evolves the state along the adiabatic path that connects to the target Hamiltonian, then one is guaranteed to obtain the ground state of the target Hamiltonian according to the well-celebrated adiabatic theorem \cite{Farhi:2001,Messiah:1962}. By the consequent QPE algorithm \cite{Kitaev:1995,Abrams:1997gk,Abrams:1998pd}, the ground state energy of the target Hamiltonian can be obtained. The ASP-QPE algorithm can also be generalized to solve the excited states and their eigenenergies \cite{Liu:2020}, as we will show in this work. In view of the fact that the computational complexity of the ASP-QPE algorithm dramatically increases due to the emergence of small energy gap(s) (referred to as ``level crossing" in the text) during the adiabatic evolution, we investigate two ideas to address this issue (cf. Refs. \cite{Albash:2018,Farhi:2002,Farhi:2010}): 1) to design suitable reference Hamiltonians; and 2) to introduce additional perturbation terms. These two methods modify the adiabatic path and avoid the level crossing such that the efficiency of the ASP procedure as well as the probability for the QPE to observe the desired eigenenergy are improved. 

We illustrate the ASP-QPE algorithm by solving the deuteron ground state energy. For the illustration of solving excited states, we apply the ASP-QPE algorithm to investigate the spectrum of the deuteron bounded in an external harmonic oscillator trap, where this system mimics the neutron-proton pair that exists in the mean field of the nuclear environment. 

We note that the ASP-QPE algorithm takes deep circuits with extra ancilla qubits, which makes it hard to execute the algorithm in near-term quantum platforms \cite{Preskill:2018}. However, since the qubitization procedure is simple (we adopt a compact mapping scheme to map the Hamiltonian eigenvalue problem onto the quantum computer) and the ASP prepares deterministically the eigenstate of interest from the input reference state, one can generalize the ASP-QPE to complicated nuclei directly, without the complexity of designing the ansatz trial function and solving high-dimensional optimization problems as in the VQE approach. Also, since the target Hamiltonians are large sparse matrices for complex nuclei, efficient quantum circuits can be designed for the simulation \cite{Aharonov:2003,Jordan:2009}: with the level-crossing problem addressed here, the efficiency of the ASP-QPE algorithm can be improved, and we expect that the algorithm will provide a promising candidate for {\it ab initio} nuclear structure calculations on future quantum computers.

This paper is organized as follows. We present the theory in Sec. \ref{sec:theory}, where we discuss various target Hamiltonians, our designs of the reference Hamiltonians and the adiabatic path. In Sec. \ref{sec:algorithm}, we discuss the ASP-QPE algorithm, where we also present our ideas to avoid the level crossing. We show the simulation conditions in Sec. \ref{sec:simulationConditions} of the model problem, where we also discuss the compact mapping scheme, the preparation of the reference states, and the design of the quantum circuit. We present the results and discussions in Sec. \ref{sec:resultsAndDiscussions}. We conclude in Sec. \ref{sec:summaryAndOutlook}, where we also present future plans.

\section{Theory}
\label{sec:theory}

\begin{table}[H]
\centering
\caption{The summary of the target Hamiltonians, the reference Hamiltonians, and the adiabatic paths in this work. See the text for details.}
\begin{tabular}{c c c}
\hline \hline 
                         &  {Natural deuteron}                    & {Trapped deuteron}	 \\ 
\hline 
{Target Hamiltonian}     & $H_{\rm target,1}= T_{\rm rel} + V_{\rm NN}$ & $H_{\rm target,2}= T_{\rm rel} + V_{\rm NN} + {\alpha [V_{\rm HO} + V_{\rm depth}] }$  \\ 
{Reference Hamiltonian}  & $H_{\rm ref,1} = H_{\rm HO} + V^{(0)}_{\rm NN} + V_{\rm shift} $  & $H_{\rm ref,2} = H_{\rm HO} + V^{(0)}_{\rm NN} + \alpha V_{\rm depth} + V_{\rm shift} $ \\ 
{Adiabatic path}         & $ H_1(t) = [1-f(t)]H_{\rm ref,1} + f(t)H_{\rm target,1} $                  & $ H_2(t) = [1-f(t)]H_{\rm ref,2} + f(t)H_{\rm target,2}  $ \\
{Path modification}      & {N.A.}       & $ H'_2(t) = H_2(t) + h_{i,j}(t) [1- h_{i,j}(t)] H^{i,j}_{\rm path} $ \\
\hline \hline 
\end{tabular} 
\label{tab:summary}
\end{table}

In this work, we apply the ASP-QPE algorithm to solve for the ground state energy of the deuteron. We also generalize the method to solve the low-lying spectrum of simple nuclear systems: we illustrate this exploration with the trapped deuteron (deuteron system bounded in an external 3DHO trap). In order to improve the efficiency of the algorithm, we discuss the ideas for 1) adopting proper reference Hamiltonian for the ASP algorithm; and 2) modifying the adiabatic path by additional perturbation terms. In Table \ref{tab:summary}, we summarize the Hamiltonians and adiabatic paths applied in this work, where the detailed discussions are presented in the following subsections.

\subsection{The natural deuteron}
We start from the intrinsic Hamiltonian of the deuteron: 
\begin{align}
{H}_{\rm target,1} = {T}_{\rm rel} + {V}_{\rm NN} \label{eq:targetHamiltonianNaturalDeuteron}, 
\end{align}
where ${T}_{\rm rel}$ denotes the relative kinetic energy of the neutron-proton system and $V_{\rm NN} $ is the nuclear interaction. In general, $V_{\rm NN} $ can be taken from various model-dependent nuclear interactions (see, e.g., Refs. \cite{Machleidt:1989,Machleidt:1987}), or those derived from first principles, e.g., chiral effective field theory (see, e.g., Refs. \cite{Machleidt:2003,Epelbaum:2008ga,Epelbaum:2014efa}). 

We follow Ref. \cite{Barrett:2013nh,Navratil:2000ww,Navratil:2000gs,Vary:2018jxg} and use the three-dimensional harmonic oscillator (3DHO) basis throughout this work. We set the oscillator energy $\hbar \omega = 5$ MeV. For the deuteron, we take the orbital angular momenta $l= 0, \ 2$, the total spin $S=1$, the  total angular momentum (coupled from the orbital angular momentum and the total spin) $J=1$, and the magnetic projection $M=0$. Together with the principle quantum number $n$, we have the complete set of quantum numbers for expressing the 3DHO basis for the deuteron structure calculation \cite{Barrett:2013nh,Navratil:2000ww,Navratil:2000gs,Vary:2018jxg}. The matrix element for the kinetic energy reads \cite{Du:2018tce}:
\begin{align}
   & \langle n' l' S' J' M' | {T}_{\rm rel} | nlSJM \rangle \nonumber \\
 = & \frac{\hbar \omega}{2} \delta ^{l'}_{l} \delta ^{S'}_{S} \delta ^{J'}_{J} \delta ^{M'}_{M} \Big[    
(2n' + l' + \frac{3}{2}) \delta _n^{n'} + \sqrt{(n'+l'+\frac{3}{2})(n'+1)} \delta ^{n'} _{n-1} + \sqrt{(n + l +\frac{3}{2})(n +1)} \delta ^{n'} _{n+1}
\Big]  .
\end{align}

It is also convenient to introduce the harmonic oscillator potential ${V}_{\rm HO} $, which admits the following analytic form \cite{Du:2018tce}:
\begin{align}
   & \langle n' l' S' J' M' | {V}_{\rm HO} | nlSJM \rangle \nonumber \\
 = & \frac{\hbar \omega}{2} \delta ^{l'}_{l} \delta ^{S'}_{S} \delta ^{J'}_{J} \delta ^{M'}_{M} \Big[    
(2n' + l' + \frac{3}{2}) \delta _n^{n'} - \sqrt{(n'+l'+\frac{3}{2})(n'+1)} \delta ^{n'} _{n-1} - \sqrt{(n + l +\frac{3}{2})(n +1)} \delta ^{n'} _{n+1}
\Big]  .
\end{align}  
The Hamiltonian of the 3DHO then reads:
\begin{align}
H_{\rm HO} = {T}_{\rm rel} + {V}_{\rm HO},
\end{align}
and the corresponding matrix element is 
\begin{align}
\langle n' l' S' J' M' | {H}_{\rm HO} | nlSJM \rangle = (2n' + l' + \frac{3}{2}) \hbar \omega   \delta _n^{n'} \delta ^{l'}_{l} \delta ^{S'}_{S} \delta ^{J'}_{J} \delta ^{M'}_{M} .
\end{align}
We note that both the ${T}_{\rm rel} $ and $ {V}_{\rm HO} $ matrices are of the tri-diagonal form (sparse matrices), while the $ H_{\rm HO} $ matrix is diagonal in the 3DHO basis. 

Finding a suitable choice of the reference Hamiltonians is important for improving the computational efficiency of the ASP-QPE algorithm (cf. Ref. \cite{Albash:2018} and references therein). We design a reference Hamiltonian $H_{\rm ref,1}$ to be our first approximation to the target Hamiltonian $ H_{\rm target,1} $. For the case of the natural deuteron, we model the reference Hamiltonian  $H_{\rm ref,1}$  as a harmonic oscillator (with its mass being the reduced mass of the two-nucleon system) present in a mean nuclear field ${V}^{(0)}_{\rm NN}$:
\begin{align}
H_{\rm ref,1} = {H}_{\rm HO} + {V}^{(0)}_{\rm NN} + {V}_{\rm shift} \label{eq:referenceHamiltonianNaturalDeuteron} ,
\end{align}
where ${V}^{(0)}_{\rm NN}$ is taken to be the diagonal entries of $V_{\rm NN}$, which serves as our approximation to the role of the mean nuclear field. Note that Hartree-Fock calculations of nuclear structure theory \cite{Delaroche:2009fa,Suhonen:2007} are conveniently used to provide the mean field. To lift the possible (near-)degeneracy of the reference Hamiltonian, we also introduce a diagonal shift Hamiltonian ${V}_{\rm shift} $ into $H_{\rm ref,1}$, which will be defined below. 

An adiabatic path connects $H_{\rm ref,1}$ to $H_{\rm target,1}$. We design the adiabatic path as:
\begin{align}
{H}_1(t) = [1- f(t)] {H}_{\rm ref,1} + f(t) {H}_{\rm target,1}  \label{eq:adiabaticPathNaturalDeuteron} ,
\end{align}
where the scalar function $f(t) \in [0,1]$ is defined within the evolution time duration $t\in [t_i,t_f]$ and it parameterizes the path. For simplicity, we take 
\begin{align}
f(t)=\frac{t-t_i}{t_f-t_i} \label{eq:timeVaryingFunction},
\end{align}
while other designs of the adiabatic path and choices of the scalar function are also applicable (cf. Ref. \cite{Schiffer:2021}). It is clear that 
\begin{align}
{H}_1(t_i) = {H}_{\rm ref,1}, \ {H}_1(t_f) =  {H}_{\rm target,1}  . 
\end{align} 
It is worth noting that we select ${H}_{\rm ref,1} $ to be diagonal. The corresponding eigenstates are the standard bases in the 3DHO representation (see Eq. \eqref{eq:standardBasesIn3DHO} below for an example). These eigenstates are taken as the reference states, which are input to the ASP algorithm.

\subsection{The trapped deuteron}
\label{sec:theTrappedDeuteron}
In order to show our idea of solving the low-lying spectrum of the nuclear system (including both the ground and excited states) and that of improving the efficiency of the ASP-QPE algorithm, we also work on the model problem in which the deuteron is bounded in an external harmonic oscillator trap (note that the physical, free-space, deuteron has only one bound state). The target Hamiltonian in the relative coordinates (we factorize and henceforth omit the center-of-mass motion as illustrated in the problems in Ref. \cite{Vary:2018jxg}) is
\begin{align}
H_{\rm target,2} = T_{\rm rel} + V_{\rm NN} + {\alpha (V_{\rm HO} + V_{\rm depth}) } \label{eq:targetHamiltonianTrappedDeuteron} ,
\end{align}
where $V_{\rm trap} = \alpha (V_{\rm HO} + V_{\rm depth})$ encapsulates the role of the external trap on the relative variables with $\alpha$ defining the trap strength (set to be 0.5). $\alpha V_{\rm depth} $ is a diagonal Hamiltonian defining the depth of the trap (the diagonal entries of $\alpha V_{\rm depth} $ are constant and set to be -12 MeV).

As our first approximation to this target system, we design the reference Hamiltonian to be 
\begin{align}
H_{\rm ref,2} = H_{\rm HO} + V^{(0)}_{\rm NN} + \alpha V_{\rm depth} + V_{\rm shift}
\label{eq:referenceHamiltonianTrappedDeuteron} ,
\end{align} 
which represents a harmonic oscillator in the presence of a model for a mean field $V^{(0)}_{\rm NN} + \alpha V_{\rm depth}$. $V_{\rm shift} $ is again applied to lift the possible (near-)degeneracy of $H_{\rm ref,2}$ and, again, it will be defined below. We construct $H_{\rm ref,2} $ to be diagonal such that its eigenstates are the standard bases in the 3DHO representation.

Similar to the natural deuteron case [Eq. \eqref{eq:adiabaticPathNaturalDeuteron}], the adiabatic path for solving the eigenstates of the trapped deuteron is designed to be
\begin{align}
{H}_2(t) = [1- f(t)] {H}_{\rm ref,2} + f(t) {H}_{\rm target,2}  \label{eq:adiabaticPathTrappedDeuteron} .
\end{align}
where ${H}_2(t_i) = H_{\rm ref,2} $ and $ {H}_2(t_f) = H_{\rm target,2}$. 

We remark that the proper design of ${H}_{\rm ref,2}$ is important to avoid the level crossing(s) during that adiabatic evolution (cf. Ref. \cite{Albash:2018} and references therein): we find that it would be desirable to include also the diagonal entries of $ V_{\rm trap} $ in ${H}_{\rm ref,2}$ such that no crossing exists. However, we deliberately exclude these entries in Eq. \eqref{eq:referenceHamiltonianTrappedDeuteron} to retain the level crossing, such that we can illustrate the alternative idea to treat the crossing problem.

This idea originates with an insight from many-body theory. In particular, given that the crossing occurs between the $i^{\rm th}$ and $j^{\rm th}$ levels\footnote{The $i^{\rm th}$ and $j^{\rm th}$ levels are the neighboring levels and we take $j=i+1$.} of the instantaneous Hamiltonian ${H}_2(t)$ during the period $ [t_a, t_b] $, we propose to introduce an extra perturbation term to modify the adiabatic path ${H}_2(t)$ for this period. Serving as a temporary perturbation, we require this term to strengthen the level interplay between the $i^{\rm th}$ and $j^{\rm th}$ levels, such that the crossing is avoided. As such, we propose the following form of the perturbation term:
\begin{align}
P^{i,j}_{\rm path}(t) = h_{i,j}(t) [1- h_{i,j}(t)] {H}^{i,j}_{\rm path} \label{eq:pathModification}.
\end{align}
The modified adiabatic path is
\begin{align}
H'_2(t) = H_2(t) + h_{i,j}(t) [1- h_{i,j}(t)] H^{i,j}_{\rm path}  \label{eq:ModifiedAdiabaticPath} . 
\end{align}
We require that $P^{i,j}_{\rm path}(t)$ varies slowly and smoothly within $ [t_a, t_b]$, and it vanishes when $t \le t_a $ and $t\ge t_b$. For simplicity, we choose $ h_{i,j}(t)={(t-t_{a})}/{(t_b-t_a)} $, with $t\in [t_a, t_b]$. The term $h_{i,j}(t) [1- h_{i,j}(t)]$ then peaks at $t= (t_a + t_b)/2 $ with the maximum $1/4$. In particular, we have constructed the $ H^{i,j}_{\rm path} $ to be of non-zero matrix elements only in the subspace $\mathcal{S}$ spanned by the reference states that later cross. Recall we design $H_{\rm ref,2} $ to be diagonal, and the corresponding reference states (eigenstates of $H_{\rm ref,2} $) are the standard bases. Therefore, the subspace $\mathcal{S}$, and the corresponding submatrix, can be easily determined. For simplicity, we take only the off-diagonal elements in the submatrix to be non-zero, while setting the diagonal elements to be zeros. For the remaining entries of ${H}^{i,j}_{\rm path}$, we set them to be zeros (see in Sec. \ref{sec:trappedDeuteronResults} for a detailed example). 

The impact of $P^{i,j}_{\rm path}(t) $ on the $i^{\rm th}$ and $j^{\rm th}$ levels is at the second order of perturbation, while its influence on the rest of the spectrum will be of higher order. We also note that additional path modifications of the form $P^{i,j}_{\rm path}(t)$ would be necessary when there are multiple level crossings. We comment that this method will likely fail when the spectrum (or subspectrum) of the target Hamiltonian is near-degenerate, since it is not practical to insert the perturbation terms (to modify the adiabatic path) at the end of the evolution.

\section{The algorithm}
\label{sec:algorithm}
In this section, we show the methodology to compute the deuteron ground state energy via the ASP-QPE algorithm on the quantum computer. The calculation of the spectrum of the trapped deuteron is analogous. We also discuss on the application of the path-modification procedure to improve the computational efficiency. 

\subsection{The deuteron ground state energy}
\begin{figure}[H]
\centering
\includegraphics[width=10cm]{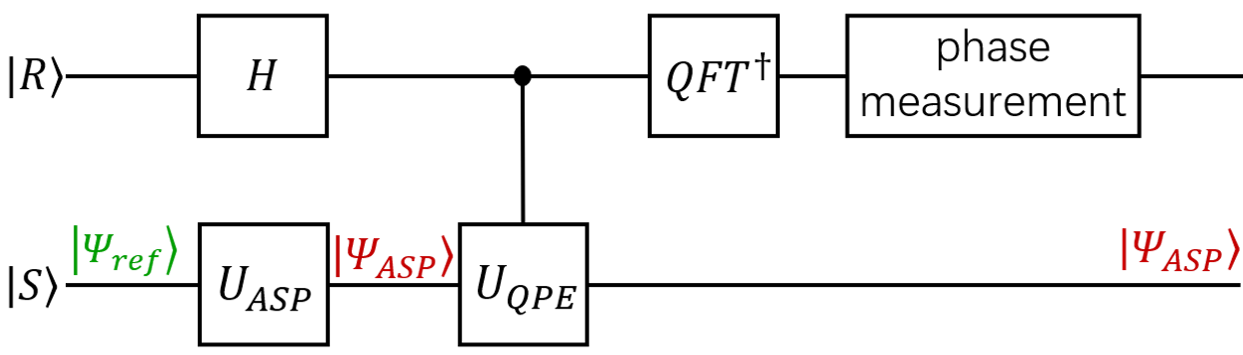}
\caption{(Color online) Illustration of the ASP-QPE algorithm. $|\Psi _{\rm ref} \rangle $ denotes the reference state, which is typically chosen as the eigenstate of the reference Hamiltonian. $|\Psi _{\rm ASP} \rangle $ is the trial eigenstate prepared by the ASP algorithm, which is then input to the QPE algorithm to evaluate the corresponding eigenenergy.}
\label{fig:IllustrationCircuit}
\end{figure}

As illustrated in Fig. \ref{fig:IllustrationCircuit}, the calculation of the deuteron ground state energy is accomplished in three steps: 1) preparing the deuteron ground state by the ASP scheme with the qubit(s) in the quantum register $S$ (state); 2) applying the QPE algorithm to encode the eigenenergy into the relative phase shift on the probing qubit(s) in the register $R$ (readout); and 3) applying the inverse Quantum Fourier Transformation ($QFT^{\dag}$) \cite{Nielsen:2001} to access the phase shift and, in turn, the eigenenergy.

In the first step, we apply ASP to prepare the trial eigenstate of the deuteron. To make the ASP algorithm efficient, proper choice of the reference Hamiltonian is crucial. In particular, one wants 1) to make the reference state easy to prepare (we design the reference Hamiltonian to be diagonal such that the reference states are the standard bases); 2) to avoid closely spacing reference states (we introduce the shift Hamiltonian to lift the (near-)degeneracy); and 3) to design the reference Hamiltonian such that it resembles the features of the target Hamiltonian (e.g., in quantum chemistry, it is advantageous to apply the Hartree-Fock Hamiltonian as the approximation to the full CI Hamiltonian \cite{Whitfield:2011}; we model our reference Hamiltonian as the harmonic oscillator as an approximation to the role of the mean nuclear field that arises from the other nucleons).

We adiabatically evolve the ground state of $H_{\rm ref,1}$ along the path [Eq. \eqref{eq:adiabaticPathNaturalDeuteron}]. According to the adiabatic theorem \cite{Albash:2018,Messiah:1962,Farhi:2001}, we are guaranteed to obtain the deuteron ground state at the end of the evolution. This adiabatic evolution can be solved as \cite{Wu:2002,Wu:2020,Schiffer:2021,Du:2020glq}: 
\begin{align}
|\psi ;t_f \rangle = {U}(t_f;t_i) |\psi ;t_i \rangle = \hat{T} \Bigg\{ \exp \Big[ -i \int _{t_i}^{t_f} {H}_1(t') dt' \Big] \Bigg\} |\psi ; t_i \rangle \label{eq:solutionOFSchrodingerEq} ,
\end{align}
where ${U}(t_f;t_i) $ is the time-evolution operator that propagates the system from $t_i$ to $t_f$, while $\hat{T}$ is the time-ordering operator. $|\psi ;t_i \rangle = |\Psi _{\rm ref} \rangle $ denotes the reference state (input), while $ |\psi ;t_f \rangle = |\Psi _{\rm ASP} \rangle $ is the output of the ASP algorithm (i.e., the trial eigenstate of the deuteron).

We can approximate the complete adiabatic evolution by a sequence of discrete adiabatic steps (time discretization). That is, we approximate the time-evolution operator $ {U}(t_f;t_i) $ as a product of $n$ unitaries \cite{Wu:2002,Wu:2020,Schiffer:2021,Du:2020glq}:
\begin{align}
{U}(t_f;t_i) \approx {U}(t_n; t_{n-1}) {U}(t_{n-1};t_{n-2}) \cdots {U}(t_2; t_1) {U}(t_1; t_0) \label{eq:decompositionOfEvolutionUnitary} .
\end{align}
The step length $\delta t = (t_f-t_i)/n$ is set to be small such that $ || H_1(t) || \cdot \delta t  \ll 1$ (we take $\delta t = 0.01$ MeV$^{-1}$ in this work). For consistency, we set $t_n =t_f$ and $t_0 = t_i$. The unitary $ {U}(t_k; t_{k-1})$ for the adiabatic step at $t_k$ can be further approximated by the first-order Trotter decomposition \cite{Nielsen:2001} based on the adiabatic path [Eq. \eqref{eq:adiabaticPathNaturalDeuteron}]:
\begin{align}
{U}(t_k; t_{k-1}) = e^{-i {H}_1(t_k) \delta t } = e^{ -i [1-f(t_k)] {H}_{\rm ref,1} \delta t } e^{ -i f(t_k) H_{\rm target,1} \delta t } + \mathcal{O}(\delta t^2). \label{eq:TrotterDecomposition}
\end{align}

In the second step, we apply the QPE algorithm \cite{Nielsen:2001} to the trial ground state $|\Psi _{\rm ASP} \rangle $, which, presumably, overlaps significantly with the exact deuteron ground state. In practice, we construct the unitary operator $U_{\rm QPE}= e^{i 2 \pi \frac{H_{np}+ \varphi }{ \tau}}$ to encode the deuteron ground state energy $E$ into the relative phase $\phi$:
\begin{align}
{U}_{\rm QPE} | \Psi _{\rm ASP} \rangle = e^{i 2 \pi \frac{H_{\rm target,1} +\varphi }{ \tau}} | \Psi _{\rm ASP} \rangle = e^{i 2 \pi \frac{E +\varphi }{ \tau}} | \Psi _{\rm ASP} \rangle = e^{i 2\pi \phi } | \Psi _{\rm ASP} \rangle .
\end{align}
Through repeated controlled actions of powers of $ {U}_{\rm QPE} $, the relative phase is encoded into the bit-string value on the qubits in the $R $ register. We note that the iterative phase estimation algorithm \cite{Aspuru-Guzik:2005} can improve the precision using small number of qubits in the $R$ register.

In the last step, we apply the inverse quantum Fourier transformation (QFT$^{\dag}$) \cite{Nielsen:2001} to the $R $ register to obtain an approximation of $\phi $ written to $R$ in binary. By classical approximation methods, we obtain an initial estimation of the ground state energy as $ E = \tau \phi - \varphi $. In practice, we choose $ \tau $ and $\varphi $ based on the initial estimation of $E$ such that the relative phase is $ 0 \leq \phi (\approx \frac{1}{2}) < 1 $. 

\subsection{The trapped deuteron and the modified adiabatic path}

In principle, the ASP-QPE approach can also be generalized to solve the low-lying spectrum of the nuclear systems as well \cite{Liu:2020}. Analogous to solving the deuteron ground state energy, we calculate the low-lying spectrum of $ H_{\rm target,2} $, i.e., the deuteron bounded in a harmonic oscillator trap. In particular, one prepares the trial eigenstates (both the bound and the excited states) of the trapped deuteron by adiabatically evolving respective reference states (i.e., the eigenstates of the reference Hamiltonian $H_{\rm ref,2} $) along the adiabatic path $H_2(t)$. With the subsequent implementation of the QPE algorithm, the eigenenergies of the low-lying states can be obtained.

We note that the level-crossing problem poses challenges to the ASP-QPE algorithm \cite{Whitfield:2011,Farhi:2001,Farhi:2002,Farhi:2010,Dickson:2011} in solving the spectrum efficiently. While the proper design of the reference Hamiltonian is helpful, we can also modify the adiabatic path by introducing perturbation terms to avoid the crossing. As discussed in Sec. \ref{sec:theTrappedDeuteron}, such modifications are only effective during the period of level crossings. They strengthen the interplay between the crossing levels such that the level crossings are avoided \cite{Tannoudji}. The efficiency of the ASP-QPE algorithm can then be improved. We illustrate this idea by solving the spectrum of the trapped deuteron, where we insert one additional perturbation term to the adiabatic path $H_2(t)$ in order to avoid the level crossing.

\section{Simulation conditions}
\label{sec:simulationConditions}
In this exploratory work, we retain a limited 4-dimensional model space for the deuteron channel. For this restricted model space, the principle and angular momentum quantum numbers of the 3DHO bases are $nl= 0s,\ 1s, \ 0d,\ 2s $.\footnote{We follow the spectroscopic notation in Ref. \cite{Suhonen:2007}. The lower case Roman letters are conventional for basis states: the quantum number of the angular momentum $l=0$ is denoted by $s$, while $l=2$ is denoted by the $d$.} In order to recover the deuteron ground state energy, we take $V_{\rm NN}$ to be the effective nuclear interaction obtained from the {\it Daejeon16} inter-nucleon interaction \cite{Kim:2019gnl} via the Okubo-Lee-Suzuki similarity transformation \cite{Vary:2018jxg}:
\begin{align}
V_{\rm NN} =
\begin{pmatrix}
-5.72084 & -3.51148 & -0.164215 & -0.533112 \\ 
-3.51148 & -8.72219 & -0.0348415 & -5.73651 \\ 
-0.164215 & -0.0348415 & -8.66007 & -0.0436011 \\ 
-0.533112 & -5.736510 & -0.0436011 & -13.8506
\end{pmatrix} .
\end{align}
Note that, unlike the effective interaction above, the nuclear interaction for the applications of complicated nuclei are large and sparse matrices \cite{Vary:2018hdv}. To lift the (near-)degeneracy of the reference states, we introduce the diagonal matrix $V_{\rm shift}$ with entries to be $+3$ MeV ($-3$ MeV) on the first (second) diagonal entry and alternating again for the third and fourth diagonal entries.

With the setup in this work, the eigenstates of the diagonal reference Hamiltonians, $H_{\rm ref,1}$ and $H_{\rm ref,2}$, are the standard bases in the 3DHO representation:
\begin{align}
|1 \rangle = (1,0,0,0)^T, \ |2 \rangle = (0,1,0,0)^T, \ |3 \rangle = (0,0,0,1)^T, \ |4 \rangle = (0,0,1,0)^T . \label{eq:standardBasesIn3DHO}
\end{align}
The ordering of basis states is, in principle, arbitrary and our chosen ordering does not produce ordered diagonal entries.  
Another choice of ordering should not affect the results of this work.

Analogous to the compact encoding scheme shown in Ref. \cite{Du:2020glq}, we map the 3DHO bases to distinct qubit configurations formed by a sequence of binaries, where each binary corresponds to one qubit in the quantum register $S$. In our model problem, the mapping from the standard bases (reference states) to the qubit configurations is
\begin{align}
|1 \rangle \mapsto |00 \rangle, \ |2 \rangle \mapsto |01 \rangle ,\ |3 \rangle \mapsto |11 \rangle, \ |4 \rangle \mapsto |10 \rangle . 
\end{align}
With this mapping scheme, the reference states can be easily prepared on the $S$ register by the action of the Pauli-X gate(s). For more complicated applications (e.g., heavier nuclei), it takes only $ \lceil \log _2 N_{d} \rceil $ qubits to encode the information of $N_d$-dimensional model space. 

The quantum circuit of the ASP algorithm is designed following Ref. \cite{Du:2020glq}. In particular, the unitary $ {U}(t_k; t_{k-1}) $ of a discretized adiabatic step in Eq. \eqref{eq:decompositionOfEvolutionUnitary} is converted into a series of unitaries: some are time-invariant unitaries that apply in other adiabatic steps, while others are diagonal unitaries with entries being explicit functions of $t_k$. We adopt standard procedures to decompose each unitary into quantum gates \cite{Li:2012,Nielsen:2001}. Combining these gates in order, we obtain the circuit of $ {U}(t_k; t_{k-1}) $, which is of fixed structure with gate parameters directly parameterized by $t_k$. Indeed, the circuit of $ {U}(t_k; t_{k-1}) $ serves as an elementary module: the complete circuit for the ASP algorithm can be automatically constructed by sequentially combining these modules (with the parameter $t_k$ updated in each module) on the quantum computer. 

With the input reference state in the $S$ register, the corresponding trial eigenstate of the target Hamiltonian can be obtained by the operation of the circuit. This trial eigenstate is then input to the QPE algorithm to evaluate the eigenenergy, where the circuit of the QPE algorithm is designed based on the target Hamiltonian following standard techniques \cite{Li:2012,Nielsen:2001}. We can also interrupt the adiabatic evolution at intermediate time and design the QPE part of the circuit based on the instantaneous Hamiltonian. In this way, we can prepare the trial eigenstates of the instantaneous Hamiltonian and measure the corresponding eigenenergies during the evolution.

We remark that we do not seek optimal design of the circuit in this initial work. Our foci are 1) to explore the feasibility of applying the ASP-QPE algorithm to solve the low-lying spectrum of nuclear systems; and 2) to explore methods to improve the efficiency of the algorithm. In future work, the circuit design will be improved. We implement the ASP-QPE algorithm on the IBM Qiskit quantum simulator \cite{Santos:2017,Cross:2017} and do not take into account the noise in simulations.

\section{Results and discussions}
\label{sec:resultsAndDiscussions}

\subsection{The deuteron ground state energy}

\begin{figure}[H]
\centering
\includegraphics[width=12cm]{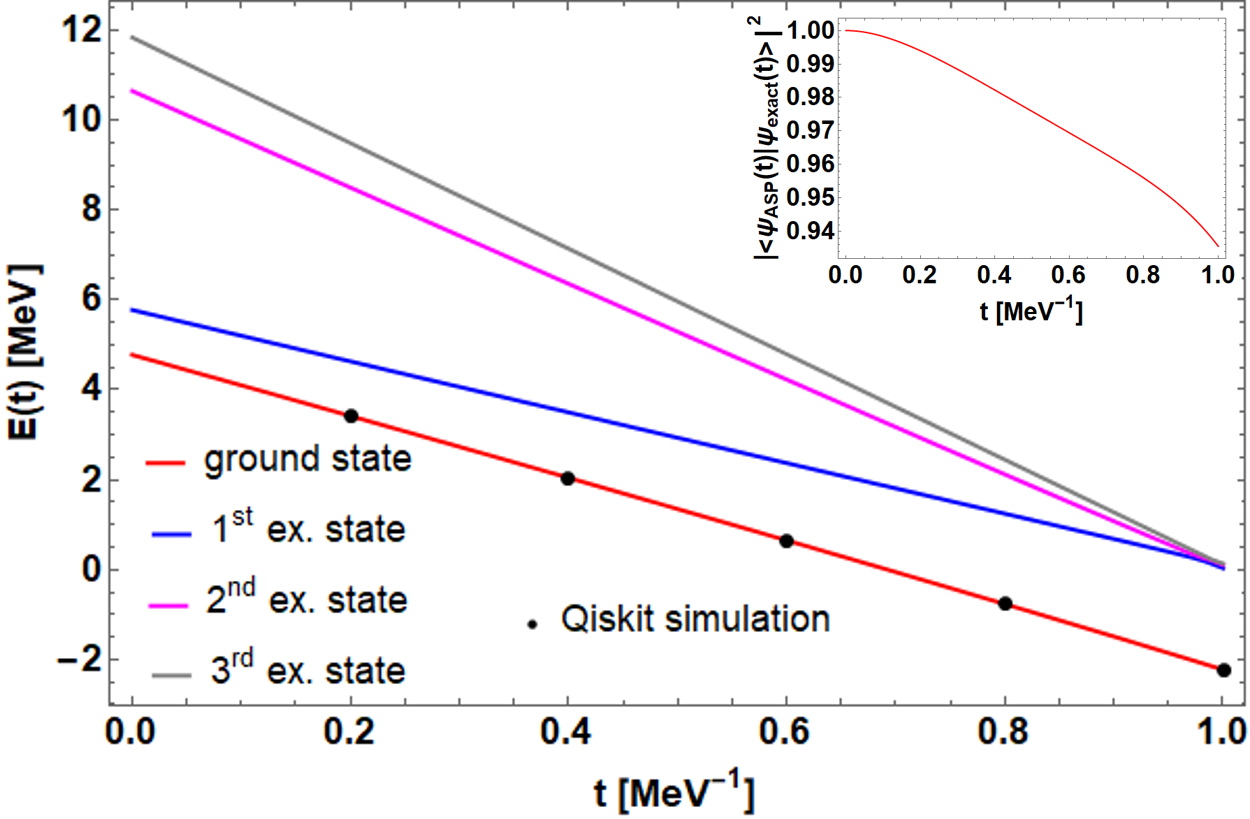}
\caption{(Color online) The evolution of eigenenergies of the instantaneous Hamiltonian $H_1(t)$. Solid lines present the results of classical calculations. The ground state energies at selected moments computed via the ASP-QPE algorithm implemented with the IBM Qiskit quantum simulator are presented as black dots. The inset panel shows the squared overlap $ | \langle \Psi _{\rm ASP} (t) | \Psi _{\rm exact}(t)  \rangle |^2 $ of the ground states of $H_1(t)$ at intermediate times during the adiabatic evolution. See the text for details.}
\label{fig:deuteronGroundState}
\end{figure}

In Fig. \ref{fig:deuteronGroundState}, we present the classical and quantum calculations of the eigenenergies of the instantaneous Hamiltonian during the adiabatic evolution. The classic results of all the four states are obtained by exact matrix diagonalization, while the quantum results are computed via the ASP-QPE algorithm for only the ground state. 

Starting from $t=0$, where the y-intercepts show the eigenvalue of $H_{\rm ref}$, we find that the energy gap between the ground state (red line) and the first excited state (blue line) is easily visible throughout the evolution. This gap implies good efficiency can be achieved when computing the deuteron ground state energy via the ASP-QPE algorithm \cite{Messiah:1962,Farhi:2001}. 

In the inset panel, we show the squared overlap $ | \langle \Psi _{\rm ASP} (t) | \Psi _{\rm exact}(t)  \rangle |^2 $ between the trial ground state, $ | \Psi _{\rm ASP} (t) \rangle  $, prepared by the ASP procedure and the exact ground state, $ | \Psi _{\rm exact} (t)  \rangle $, solved by diagonalizing the instantaneous Hamiltonian $H_1(t)$ for every adiabatic step. Both the states and the squared overlap are computed on the classical computer. The squared overlap values indicate the quality of the trial ground states at intermediate times: better quality of the trial ground states ensures higher probability for the QPE algorithm to observe the ground state energies during the evolution. We find that the ASP procedure well prepares the instantaneous ground state along the adiabatic path: the squared overlap $ | \langle \Psi _{\rm ASP} (t) | \Psi _{\rm exact}(t)  \rangle |^2 $ is above 93$\%$ throughout the evolution. We also checked that these values approach 100$\%$ given moderately longer evolution time (e.g., at $T=10$ MeV$^{-1}$, versus $T = 1$ MeV$^{-1}$ used here, the squared overlap values are all beyond 99.5$\%$). 

Implementing the IBM Qiskit quantum simulator \cite{Santos:2017,Cross:2017}, we compute the ground state energies at selected moments along the adiabatic path $H_1(t)$ based on the ASP-QPE algorithm. These results are shown as the black dots in Fig. \ref{fig:deuteronGroundState}. Note that the errors of these quantum results are mainly from the statistical variance during the measurement, which are smaller than the size of the markers. Therefore, we do not show the error bar explicitly. We find the quantum results (black dots) agree well with the classical results (red line). At $t=T=1$ MeV$^{-1}$, the deuteron ground state energy is recovered. To obtain more precise result of the deuteron ground state energy making use of limited number of qubits, a recursive phase estimation algorithm \cite{Aspuru-Guzik:2005} will be useful.

\subsection{The deuteron in the harmonic oscillator trap}
\label{sec:trappedDeuteronResults}

\begin{figure}[H]
\centering
\includegraphics[width=12cm]{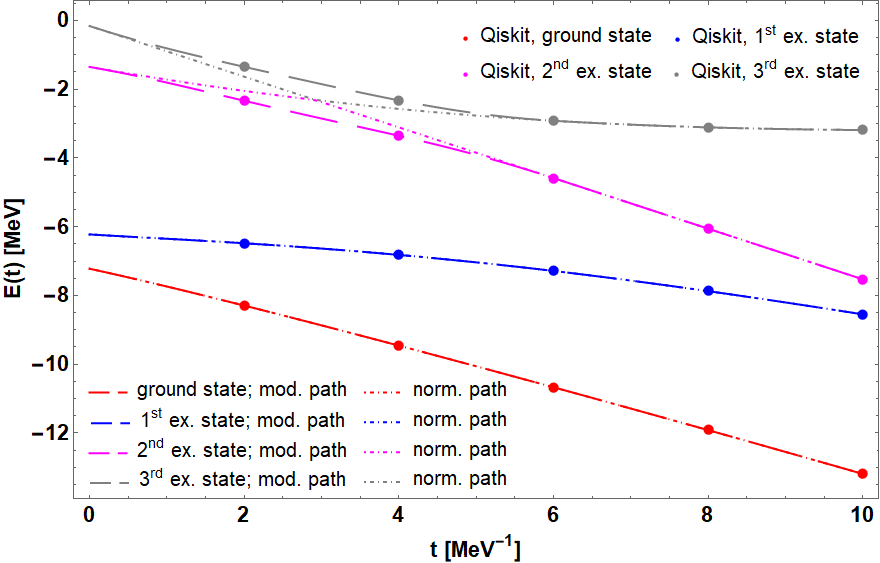}
\caption{(Color online) Evolution of the spectra via the normal adiabatic path $ {H}_{2}(t)$ (short dot-dashed lines) and via the modified adiabatic path $H'_{2}(t) $ (long dashed lines). The spectra at selected moments in time computed via the ASP-QPE algorithm with the modified adiabatic path $H_2'(t)$ implementing the IBM Qiskit quantum simulator are shown as colored dots. With the normal adiabatic path, the level crossing occurs at $t\approx 2.9 $ MeV$^{-1}$ between the $2^{nd}$ and $3^{rd}$ excited states, with the minimal gap size being $\Delta E \approx 0.02$ MeV.
}
\label{fig:onePathBoundwithTrap}
\end{figure}

The ASP-QPE algorithm can be applied to solve the eigenenergies of the excited states in addition to the ground state energy. We illustrate this scheme by solving the spectrum of the deuteron system bounded in an external harmonic oscillator trap.

We again apply the ASP algorithm to prepare the trial eigenstates of the system: each trial eigenstate (output) is evolved from the corresponding reference state (input) via the adiabatic path. The straightforward choice of the path is $H_2(t)$. By exact diagonalization of the instantaneous Hamiltonian, we compute the spectrum of $H_2(t)$ as functions of time. The results are shown as the short dot-dashed lines in Fig. \ref{fig:onePathBoundwithTrap}. Note that the y-intercepts correspond to the eigenvalues of ${H}_{\rm ref,2}$, which are indeed the diagonal entries of ${H}_{\rm ref,2}$ [Eq. \eqref{eq:referenceHamiltonianTrappedDeuteron}]. We find that the minimal energy gap between the 2$^{nd}$ and 3$^{rd}$ excited states occurs at $t_c \approx 2.9 $ MeV$^{-1}$: such a level crossing causes a transition between these two states and hinders the efficiency of the ASP algorithm \cite{Whitfield:2011,Farhi:2001,Farhi:2002,Farhi:2010,Dickson:2011,Liu:2020}.  

In order to avoid the crossing, we modify the adiabatic path. In particular, we introduce an extra perturbation term to the adiabatic path $H_2(t)$ during the period $ [t_a, t_b]$ when the levels approach (here we take $t_a=0$ and $t_b =6$ MeV$^{-1}$). We note that these crossing levels evolve from the reference states $|3 \rangle = (0,0,0,1)^T $ and $|4\rangle = (0,0,1,0)^T $, which span the subspace $\mathcal{S}=\{|3 \rangle, |4 \rangle \}$. Therefore, we design the perturbation term to be $P_{\rm path}^{3,4} = h_{3,4}(t) [1-h_{3,4}(t) ] H_{\rm path}^{3,4} $, where we take the matrix elements $[ H_{\rm path}^{3,4}]_{34}=[ H_{\rm path}^{3,4}]_{43}=-2$ MeV, while the remaining elements of the matrix $[ H_{\rm path}^{3,4}]_{4\times 4}$ are all zeros. 

The modified adiabatic path $H'_2(t)$ connects the same reference and target Hamiltonians as ${H}_{2}(t)$. We compute the spectrum of $H'_2(t)$ by exact diagonalization during the evolution. The results are shown as the long dashed lines in Fig. \ref{fig:onePathBoundwithTrap}. The additional perturbation avoids the level crossing by strengthening the interplay between between the 2$^{nd}$ and 3$^{rd}$ excited states. We remark that we have also applied this path-modification scheme to the same model problem but with an enlarged (64-dimensional) model space; we verified the feasibility of this scheme to avoid multiple level crossings during the evolution.

\begin{figure}[H]
\centering
\includegraphics[width=12cm]{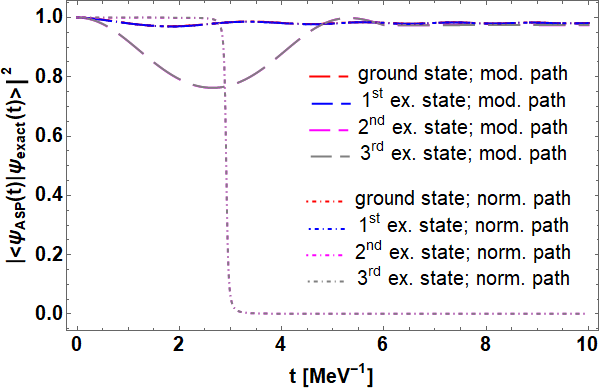}
\caption{(Color online) Evolution of the squared overlap $| \langle \Psi _{  \rm ASP} (t) | \Psi _{ \rm exact} (t) \rangle  |^2$ for each state based on the normal path $H_{2}(t)$ (short dot-dashed lines) and the modified path $H'_{2}(t)$ (long dashed lines) during the evolution. The magenta long dashed/short dot-dashed lines overlap with their respective gray lines, while all the blue lines overlap with all the red lines.}
\label{fig:graphTrapTotalOverlap}
\end{figure}

Based on either the normal path $H_{2}(t)$ or the modified path $H'_{2}(t)$, we compute for every intermediate moment $t$: 1) the $k^{th}$ ($k=1,\ 2,\ 3,\ 4$) trial eigenstate $| \Psi _{  \rm ASP;k} (t) \rangle $ evolved from the $k^{th}$ reference state by the ASP procedure; and 2) the $k^{th}$ exact eigenstate $| \Psi _{ \rm exact;k} (t) \rangle  |^2 $ obtained by matrix diagonalization. We then compute the squared overlap $| \langle \Psi _{  \rm ASP; k} (t) | \Psi _{ \rm exact ; k} (t) \rangle  |^2$ for each state. Note that these calculations are all done on a classical computer. These values characterize the quality of the trial eigenstates prepared by the ASP algorithm. 

We show the squared overlap values in Fig. \ref{fig:graphTrapTotalOverlap}. For the evolution via the normal path, we note that the squared overlap values of the ground state and of the $1^{st}$ excited state approach 1. However, the squared overlap values of $2^{nd}$ and $3^{rd}$ excited states drop to about zero starting from the crossing point at $t\approx 2.9 $ MeV$^{-1}$ (seen from the overlapping magenta and gray dot-dashed lines): a significant crossover transition takes place between these two states with the limited total evolution time. The small energy gap and the resulting transition hinder the efficiency of the ASP-QPE algorithm. 

For comparison, we present the evolution of the squared overlap value of each state computed based on the modified path $H'_{2}(t)$ in Fig. \ref{fig:graphTrapTotalOverlap}. We find that the squared overlap values of the ground state and the $1^{st}$ excited state overlap with those computed based on the norm path $H_{2}(t)$ (they all nearly approach 1). We also note that a significant crossover transition is prevented since the level crossing is avoided: 1) the squared overlap values of the $2^{nd}$ and $3^{rd}$ excited states remain above 0.75 during the evolution; 2) the squared overlap values of these two states are restored to more than 0.97 at the end of the evolution. We also checked to see that, with longer evolution time (e.g., $T \geq 50$ MeV$^{-1}$), the squared overlap values for all the states computed based on the modified path $H'_{2}(t)$ are more than 0.99 throughout the adiabatic evolution; for achieving the same quality, it would take $T \gg \frac{1}{\Delta E ^2} \approx 2500 $ MeV$^{-1}$ \cite{Messiah:1962} for the adiabatic evolution via the normal path $H_{2}(t)$. Thus, the efficiency of the ASP algorithm can be significantly improved by the path modification scheme.

The trial eigenstate $| \Psi _{  \rm ASP;k} (t) \rangle $ prepared based on the modified adiabatic path $H'_2(t)$ is input to the QPE algorithm for evaluating the eigenenergy. For the total evolution time $T=10$ MeV$^{-1}$ applied in this work, the trial eigenstates overlap significantly with the corresponding exact eigenstates in the intermediate times and at the end of the evolution. Therefore, one observes the eigenenergies of the corresponding trial eigenstates with large probabilities. 

Implementing the IBM Qiskit quantum simulator \cite{Santos:2017,Cross:2017}, we compute the spectra of $H'_2(t)$ implementing the ASP-QPE algorithm at selected times during the evolution. The results are presented as the colored dots in Fig. \ref{fig:onePathBoundwithTrap}. We find that the results obtained via the quantum computation agree well with the classical results. We do not show the error bar (which is mainly from the statistical variance in measurement and of the size smaller than the marker size). At the end of the evolution, the ASP-QPE algorithm successfully recovers the eigenenergies of the target Hamiltonian.

Finally, we notice that it is important to know where and which neighboring levels cross in order to implement the ASP-QPE algorithm efficiently. In quantum computations, we can achieve this by energy measurement at intermediate times along the adiabatic path. While the decreasing energy gap is one of the signs of the level crossing, multiple eigenvalues observed with comparable probabilities on one trial eigenstate, which persists even with longer evolution time, will also be indicative. 

In many applications, we are interested in a few low-lying states of the nuclear system. Once the crossing(s) is (are) detected, we can implement the path-modification scheme to avoid the crossing(s) and work to improve the computational efficiency. We expect that the ASP-QPE algorithm, together with the path modification scheme, will be applicable for solving the (sub)spectra with moderately separated states. However, the algorithm will be less efficient as the spectrum becomes more dense (e.g., with near-degenerate states of the target Hamiltonian), where the energy gaps between the levels of the target Hamiltonian are small.

\section{Summary and outlook}
\label{sec:summaryAndOutlook}

Solving nuclear many-body problems using an {\it ab initio} approach is a computationally hard problem. For complex nuclei with increasing atomic number, this task will be intractable even for world-leading supercomputers. Quantum computers potentially open a promising path to address such nuclear many-body problems, and there is hence a need to develop the quantum algorithms for {\it ab initio} nuclear structure theory. In this work, we investigate the application of the quantum algorithm of the adiabatic state preparation with quantum phase estimation (ASP-QPE) in solving {\it ab initio} nuclear structure theory. 

We first apply the ASP-QPE algorithm to solve the ground state energy of the natural deuteron. We apply the ASP algorithm to prepare the trial ground state. Proper design of the reference Hamiltonian is important for the algorithm to be efficient. In this work, we model the reference Hamiltonian as a harmonic oscillator (its mass being the reduced mass of the two-nucleon system) existing in the mean nuclear field. We design the adiabatic path to connect the reference Hamiltonian and the deuteron Hamiltonian. By adiabatically evolving the ground state of the reference Hamiltonian along the path, we prepare the trial ground states of the instantaneous Hamiltonians. These trial states are input to the QPE algorithm for evaluating the instantaneous ground state energies. We adopt a compact mapping scheme to map the model problem onto the quantum register. The quantum circuit of the ASP algorithm is designed following the idea of Ref. \cite{Du:2020glq} and the circuit of the QPE algorithm is designed based on the instantaneous Hamiltonians. We implement the IBM Qiskit quantum simulator to simulate the circuits and the quantum results agree well with the classical results solved by exact matrix diagonalization. At the end of the evolution, we recover the deuteron eigenenergy.

We then generalize the ASP-QPE algorithm to solve the low-lying spectra (both the ground and excited states) of simple nuclear systems. We note that the efficiency of the algorithm is hindered if there is a small/vanishing energy gap (level crossing). While suitable design of the reference Hamiltonian is helpful to address this issue, we also introduce the idea of path modification by insertions of intermediate perturbation terms. In particular, we design these perturbation terms such that they are only effective during the periods of level crossing and require these terms to strengthen the interplay between the approaching levels. The level crossing can be avoided by this path-modification scheme, via which process we manage to improve the efficiency of the ASP-QPE algorithm.

As an illustration, we solve the spectrum of the deuteron bounded in an external harmonic oscillator trap. The application of the ASP-QPE algorithm is analogous to that of the natural deuteron, except for the modified path to avoid the level crossing. Implementing the IBM Qiskit quantum simulator, we solve the eigenenergies of the instantaneous Hamiltonians at selected times. All of the quantum results agree well with the exact results obtained by matrix diagonalization. At the end of the evolution, the spectrum of the bounded deuteron is recovered. 

In the future, we plan to develop the ASP-QPE algorithm to solve the low-lying spectra of complicated nuclei. Adopting the compact encoding scheme, it takes only $ \lceil \log _2 N_{d} \rceil $ qubits to encode the $N_d$-dimensional model problem. Meanwhile, proper design of the reference Hamiltonian and the path-modification scheme help to improve the efficiency. Since all the Hamiltonians involved in constructing the adiabatic path and in the phase estimation procedure are (or can be divided into) large sparse matrices with elements explicitly computable, an efficient quantum circuit implementing the ASP-QPE algorithm can be designed \cite{Aharonov:2003,Jordan:2009,Berry:2007} and we expect that quantum advantage can be achieved.

\section*{Acknowledgments}
This work was supported in part by the US Department of Energy (DOE) under Grants No. DE-FG02-87ER40371 and No. DE-SC0018223 (SciDAC-4/NUCLEI). A portion of the computational resources were provided by the National Energy Research Scientific Computing Center (NERSC), which is supported by the US DOE Office of Science. X.Z. and W.Z. are supported by the Strategic Priority Research Program of Chinese Academy of Sciences, Grant No. XDB34000000. X.Z. is supported by new faculty startup funding from the Institute of Modern Physics, Chinese Academy of Sciences and by Key Research Program of Frontier Sciences, CAS, Grant No. ZDBS-LY-7020. W.Z. are supported by the National Natural Science Foundation of China (Grants No. 11975282 and No. 11435014) and the 973 Program of China (Grant No. 2013CB834405). 


\end{document}